\documentclass{article} 
\usepackage{nips13submit_e,times}
\usepackage{hyperref}
\usepackage{url}
\usepackage{graphicx}
\usepackage{color}

\title{Gene Co-expression Network analysis of Lung Squamous Cell Carcinoma data}

\author{
Md-Nafiz Hamid\\
Bioinformatics and Computational Biology program\\
Department of Veterinary Microbiology and Preventive Medicine \\
Iowa State University, Ames, IA, USA\\
\texttt{nafizh@iastate.edu}  \\
}

\nipsfinalcopy 

\begin{document}

\maketitle

\begin{abstract}
We performed a gene co-expression analysis on Lung Squamous Cell Carcinoma data to find modules (groups) of genes that may highly impact the growth of these type of tumors. Additionally, we used cancer survival data to relate modules to prognostic significance in terms of survival time. Analysis on RNA-seq data revealed modules which are significant in gene enrichment analysis. Specifically, two genes - RFC4 and ECT2 - have been found to be significant in terms of survival time. We also performed a second gene co-expression analysis on a second dataset of microarray data, and many significant genes found in this analysis could also be found in the RNA-seq data implying that these genes might indeed play a crucial role in Lung Squamous Cancer. All the R code for the analysis can be found at: \url{https://github.com/nafizh/Gene_coexpression_analysis_lung_cancer}
\end{abstract}

\section{Introduction}
Lung Cancer is the event of tumors in lung where uncontrolled growth of tissues happen. Lung Cancer is mainly of 2 types - Non-Small Cell Lung Carcinoma (NSCLC) and Small Cell Lung Carcinoma (SCLC). About 85\% of lung cancer that happen are of the non-small type. NSCLC is again of mainly 2 types - lung adenocarcinoma and squamous cell lung carcinoma. Mainly people who are smokers encounter the squamous carcinoma. 

Gene co-expression analysis  is an exploratory type analysis that is done to look for genes that are highly connected in a biological network, and might therefore play crucial role in the biological processes. The analysis in this paper is a Weighted Gene Co-expression Network Analysis (WGCNA) that was introduced by Langfelder et al.[1]. The purpose of this analysis is to construct a weighted co-expression network from RNA-seq gene expression data, and find modules of highly connected genes. Consequently, we also attempt to relate these modules to prognostic significance for cancer survival time, and associate gene significance with intramoduler connectivity. Concretely, we set out to find significant genes in terms of their impact on patient survival time. 

At the time of this project (Spring 2016), to the best of our knowledge, there has not been any WGCNA (Weighted Correlation Network Analysis) analysis done on lung squamous data, and investigate genes that are significant to patient survival time. Other ways of doing this type of analysis would also involve building a network first, and then find highly connected genes. Therefore, WGCNA seemed to be the best way to perform this analysis as there already exists a highly efficient R library for this purpose[2].  

\section{Background}
After we collected RNA-seq expression data, a weighted gene co-expression network was built in this analysis. We used the Pearson correlation measure between two genes to calculate connectivity through a power function. It is absolutely essential that our network has a scale free topology as a biologically meaningful network should possess this property. From these correlation measures, a topological overlap matrix (TOM) was created so that we can take into account the number of neighbors two genes share. From this network, we find highly connected modules by doing an average linkage hierarchical clustering. These modules contain tightly connected genes on which we can conduct our next analysis. We can use the first principal component of these module to search for genes that are hub genes, and potentially the crucial players in lung cancer. 

We used the Cox proportional hazards model[3] to define the prognostic significance of each gene by its univariate cox regression p-value. We use the clinical survival data from our dataset to do this. Previously, it was claimed that high values of gene significance might imply that gene expression is a significant predictor for patient survival[4]. We searched for genes that are highly connected as well as prognostically significant. We also did gene enrichment analysis on modules that had the highest average gene significance in terms of survival rate. It should be noted here that we defined gene significance in terms of patient survival time but it can really be any biologically meaningful property. 

Other ways of doing this same analysis might involve building the adjacency matrix based on another correlation measure such as Spearman correlation, and then finding nodes in terms of betweenness centrality or closeness centrality. The clustering might be done with another type of distance such as euclidean or manhattan distance rather than the topological overlap matrix values used in our analysis. Understandably, there are many variations of the network analysis that can be done iteratively, and then we can investigate how the results differ.

\section{Methodology}
Both the RNA-seq and microarray datasets were collected from the UCSC Cancer genome browser[5]. The RNA-seq data set has the gene level transcription estimates in RSEM normalized count. The data set was log normalized beforehand, which made the normalization process much easier. It has 520 samples of 10592 genes. Clinical data including patient survival time is also present in the data set. 

For constructing the network, Pearson correlation measure was calculated between each pair of genes. That correlation was then used to measure a connection strength by the following formula -
\begin{center}
   $Connection\_strength(i, j) = |cor(i, j)|^\beta$
\end{center}

For each gene its total Connectivity is  defined as the sum of connection strengths with the other network genes. For deciding $\beta$, a linear regression model fitting index is used for different values to see how accurately a resulting network would fit a scale free topology. From Figure 1, we can see that 8 is the lowest value that maintains a high approximation on scale free topology.

\begin{figure}[h]
\begin{center}
\includegraphics[width=6in]{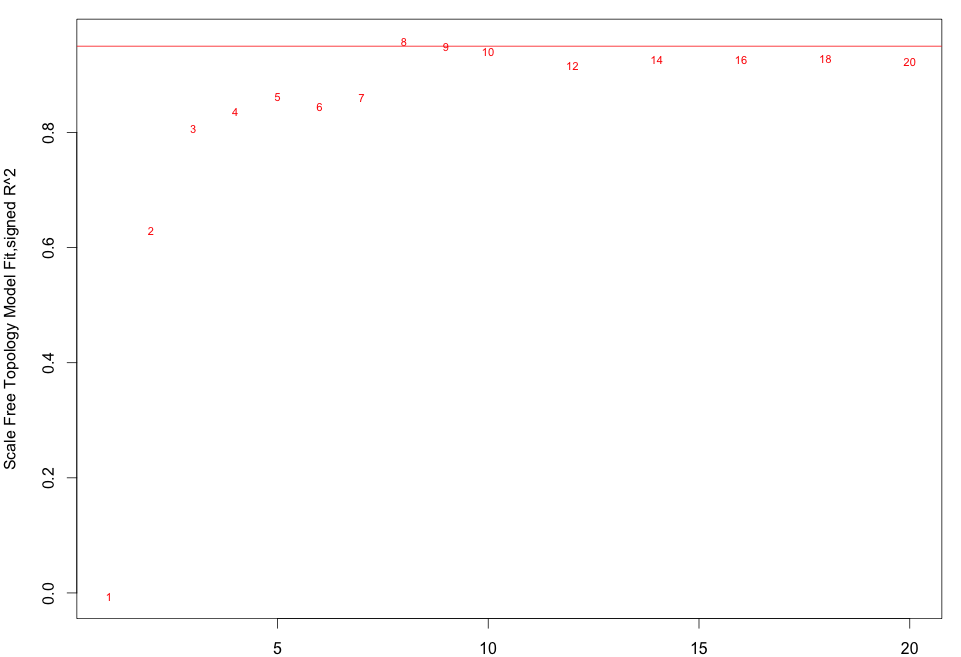}
\end{center}
\caption{Deciding on $\beta$}
\end{figure}

From the connection strengths that we calculated in the previous step, we created a Topological Overlap Matrix (TOM) where the connection between two genes $i$ and $j$ is defined as  -

\begin{center}
   $t_{ij} = \frac{|N(i) \cap N(j)| + a_{ij} + I_{i=j}}{min\{|N(i)| , |N(j)|\}+1-a_{ij}}$
\end{center}

where $N(i) =$ set of neighbors of $i$ and $I = 1$ only if $i = j$, otherwise $0$.

TOM takes into account the overlap of neighbors that two genes possess. The motivation behind using this comes from the intuition that higher overlap might indicate similar functionality.

The TOM values were then used as a distance measure to do a hierarchical clustering on 3600 most connected genes of the network. The number 3600 was decided based on trial and error, and the results do not really change if we increase the number. We then cut the cluster tree at a value of 0.94 to find our desired modules which are clusters of highly connected genes. Again, the number 0.94 was also decided by trial and error where if we increase it the results do not improve. From Figure 2, we can see that we detected 5 modules.

\begin{figure}[h]
\begin{center}
\includegraphics[width=4in]{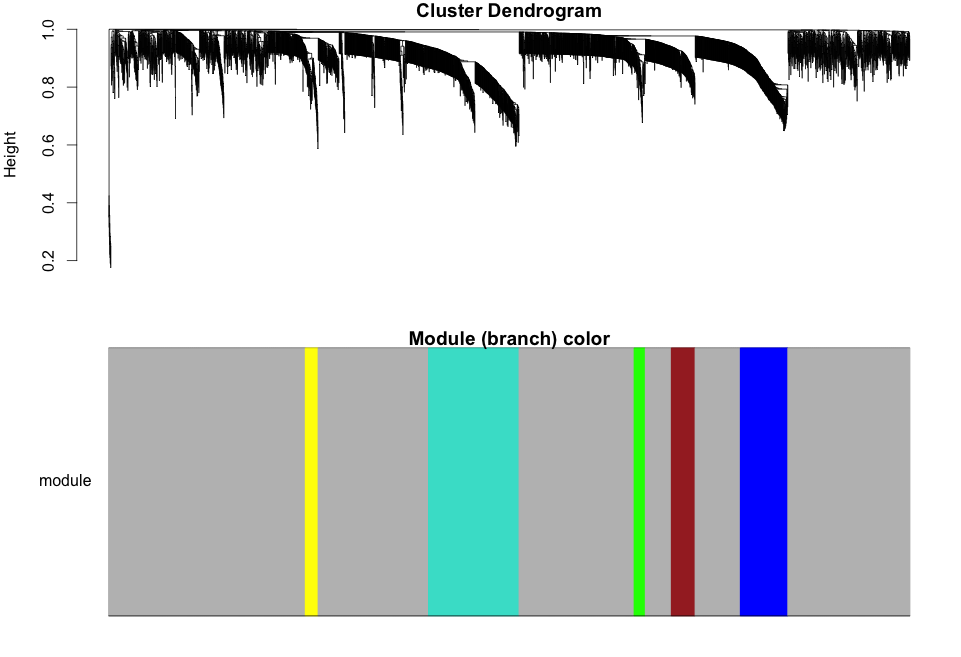}
\end{center}
\caption{Hierarchical Clustering and Module finding in RNA-seq dataset}
\end{figure}

Consequently, we calculate the gene significance for each gene in terms of the survival time of a patient. We used a survival model called cox proportional hazard model to do that. Survival models relate the time before some event occurs to one or more covariates that may be associated with that amount of time. In our case, that event is death. To define a measure of prognostic significance, we used a univariate Cox proportional hazards regression model to regress patient survival on the individual gene expression profiles. The resulting univariate Cox-regression p-values were used to define a measure of prognostic significance-

\begin{center}
   $Gene Significance = -log10(Cox pvalue)$
\end{center}

-log10 was used so that the measure of gene significance is proportional to the number of zeroes in the p-value. The resulting average gene significance in each module that was previously found can be seen below -

\begin{figure}[h]
\begin{center}
\includegraphics[width=4in]{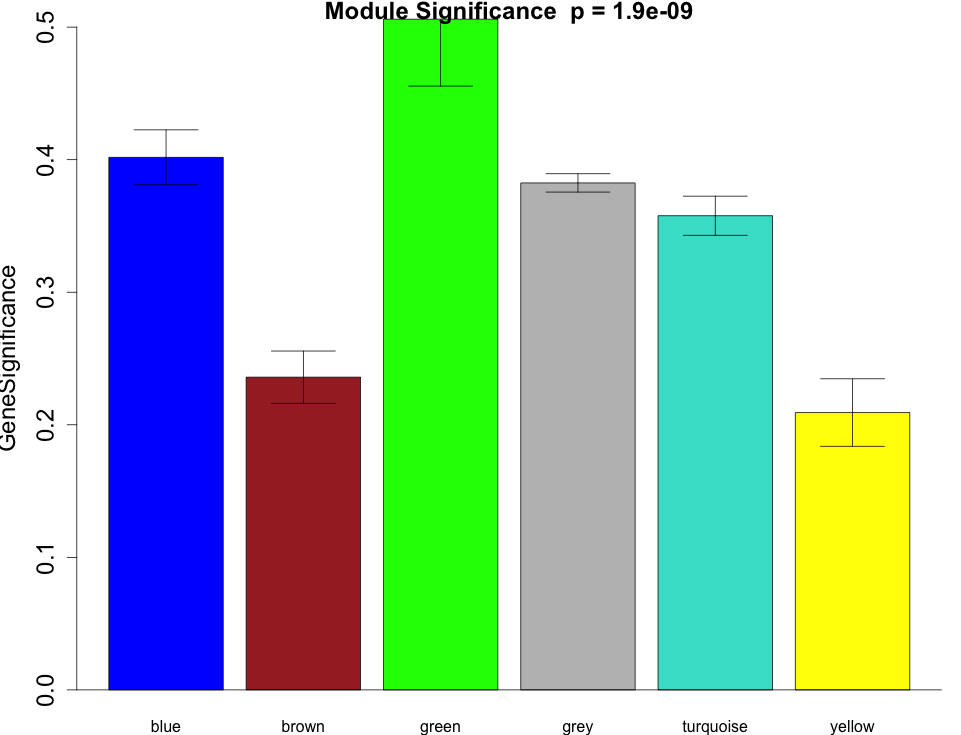}
\end{center}
\caption{Module significance in terms of Survival time}
\end{figure}

From Figure 3, we can see that the green and blue modules are the most enriched modules with significant genes. We then did a principal component analysis for each module, and to get a sense of how these modules are related we can correlate their first components with each other (Figure 4).

\begin{figure}[h]
\begin{center}
\includegraphics[width=4in]{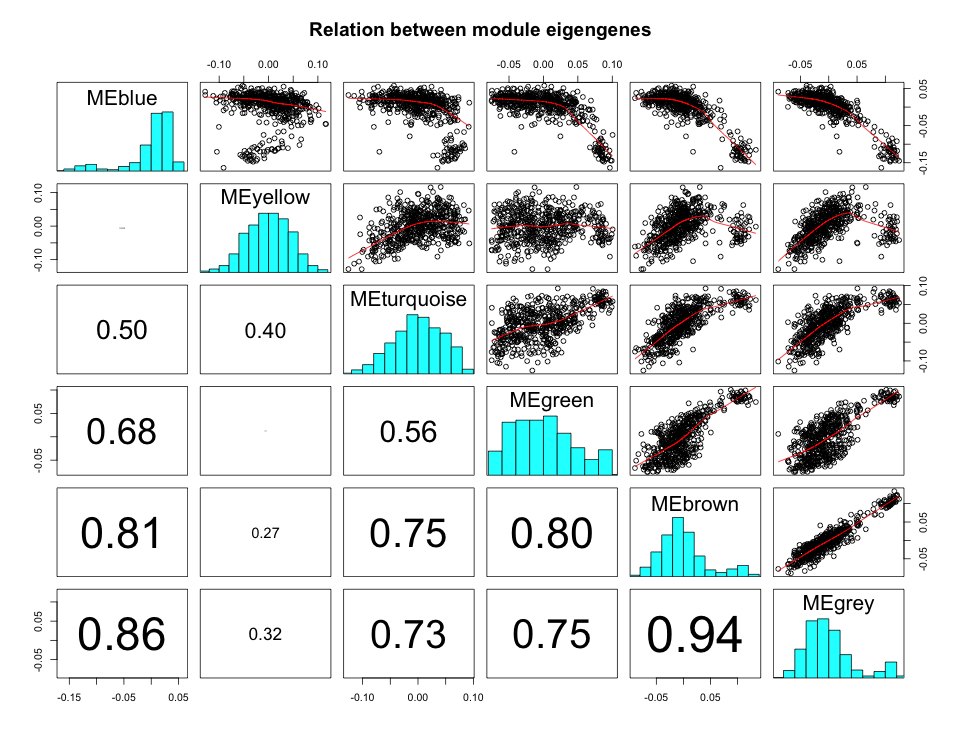}
\end{center}
\caption{Correlations between the modules}
\end{figure}

We can see from the values that the green and blue modules are also significantly correlated. By making a clustering tree from the principal components which we call the eigengenes, we can also see how they are related with each other (Figure 5).

\begin{figure}[h]
\begin{center}
\includegraphics[width=4in]{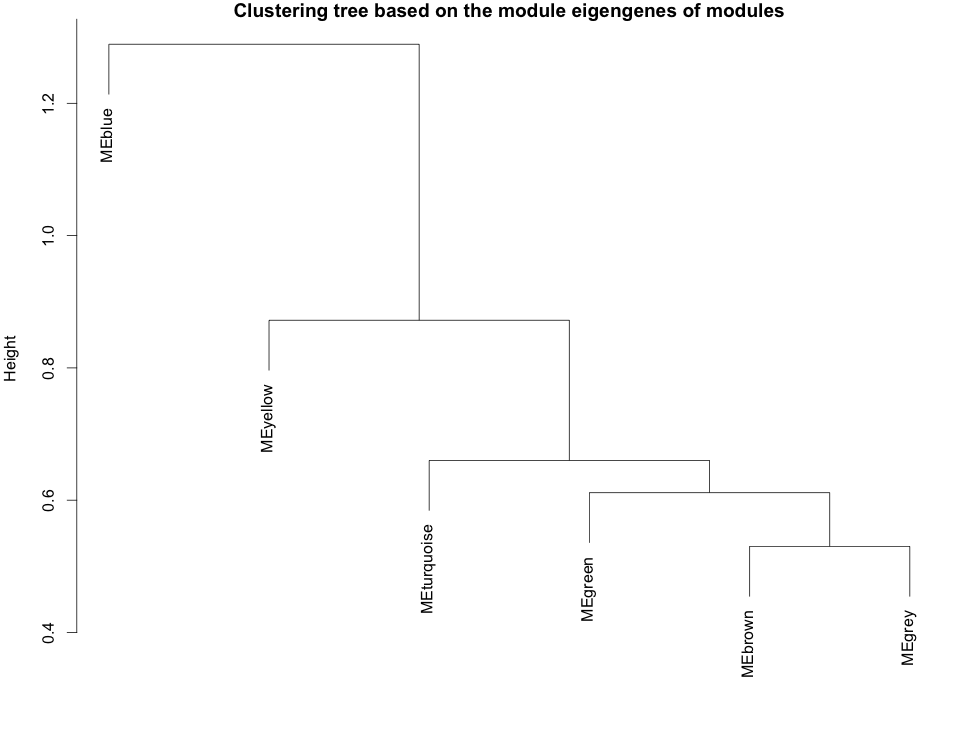}
\end{center}
\caption{Cluster tree for the Eigengenes in RNA-seq data}
\end{figure}

In the final step of our pursuit of significant genes, we calculated the intramodular connectivity for all genes in a module. For each gene, we defined a measure of module membership by correlating its gene expression profile with the module's principal component.

\begin{center}
   $K_{i}^{blue} = cor\{i, Pcomp^{blue}\}$
\end{center}

In our pursuit, we tried to find genes that have \textcolor{blue}{cox p-value $<$ 0.05} and \textcolor{blue}{intramodular\_connectivity $>$ 0.85}. We also found genes within the blue and green module that have \textcolor{blue}{intramodular\_connectivity $>$ 0.85}.

\section{Results, Analysis, Interpretation}
Among all the modules, only 2 genes were found that satisfied both conditions with high intramodular connectivity as well as a cox p-value of less than 0.05. These are RFC4 and ECT2. 

The ECT2 gene has shown to be an oncogene in human cancers[6]. ECT2 is shown to be over-expressed at the mRNA and protein levels in established NSCLC cell lines as well as primary NSCLC tumors[7]. Also, ECT2 over-expression was found in 82\% of primary NSCLC tumors. This biological finding was vindicated by our analysis which showed ECT2 gene to not only be highly connected but also significant in terms of patient survival. Moreover, ECT2 gene over-expression has shown to be associated with poor prognosis in patients with NSCLC[8]. Again, NSCLC patients whose tumors exhibit strong ECT2 staining had a poorer prognosis than patients whose tumors showed weak or no ECT2 staining which also supports our finding of ECT2 being prognostically significant.

The RFC4 gene has also shown to be a useful biomarker for aggressive breast tumors[9]. It was shown that siRNA-mediated knockdown of RFC4 significantly reduced cell proliferation in ER-negative normal breast and cancer lines. This implied RFC4 is essential for both normal and cancer cell survival. Moreover, overexpression of RFC4 was also correlated with poor survival in tumors[9] supporting our finding of RFC4 being significant in terms of survival rate.

\begin{table}[t]
\caption{Significant Genes in each module}
\label{sample-table}
\begin{center}
\begin{tabular}{ll}
\multicolumn{1}{c}{\bf Module}  &\multicolumn{1}{c}{\bf Number of Genes}
\\ \hline \\
Green         &59 \\
Blue             &134 \\
Turquoise             &222 \\
Brown           &149 \\
Yellow          &43 \\
\end{tabular}
\end{center}
\end{table}

We also found other genes that were highly connected and satisfied our intramodular connectivity condition. We did a gene enrichment analysis on the genes of the Green and Blue module as these were the top 2 significant modules for our purpose. We used David Bioinformatics Resources 6.7 web interface[10] tool to do the gene enrichment analysis. 

\begin{figure}[h]
\begin{center}
\includegraphics[width=6in]{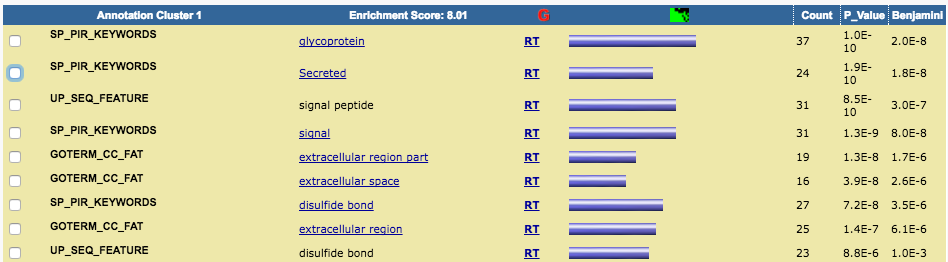}
\end{center}
\caption{Gene Enrichment Cluster on Green Module}
\end{figure}

From the analysis on green module significant genes (Figure 6), we can see that a lot of the genes are involved with glycoprotein - which, if defective, causes Bernard Soulier Syndrome (BSS), a familial coagulation disorder characterized by a prolonged bleeding time, unusually large platelets. Moreover, 19 genes were directly associated with lung cancer in disease databases. Again, 20 genes were associated with cancer/metabolic/cardiovascular/immune diseases. A lot of the genes are also connected with the GO term extracellular region/space. These information might provide crucial insight into verifying these genes role via biological experiments.

For the blue module(figure 7), we see that most of the genes are associated with cell cycle/mitotic which indicate that most of these genes might be involved in Cancer tumor growth. 

\begin{figure}[h]
\begin{center}
\includegraphics[width=6in]{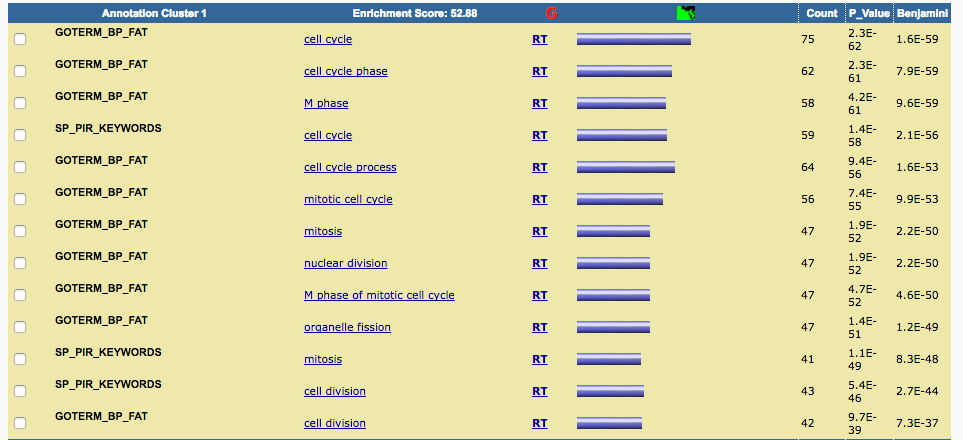}
\end{center}
\caption{Gene Enrichment Cluster on Blue Module}
\end{figure}

Moreover, 28 genes were associated with lung cancer in disease databases. Also, 113 genes were associated with phosphoprotein. Phosphoprotein secretome in tumor cell has been a source of candidates for Breast Cancer Biomarkers in Plasma[11]. Tumor cells secrete proteins into the extracellular environment. Some of these proteins could reach circulation, and become suitable biomarkers for improving diagnosis or monitoring response to treatment[11]. Also, the term \textbf{"extracellular environment"} is crucial here as we found out a significant number of genes in the green module are associated with this. 

\section{Comparison}
We wanted to verify that we can reproduce these modules, and their emergence was not a fluke. Therefore, we repeated the whole analysis on a microarray dataset with the same set of genes for 129 samples. We found 8 modules from our analysis on the microarray dataset (Figure 8).

\begin{figure}[h]
\begin{center}
\includegraphics[width=4in]{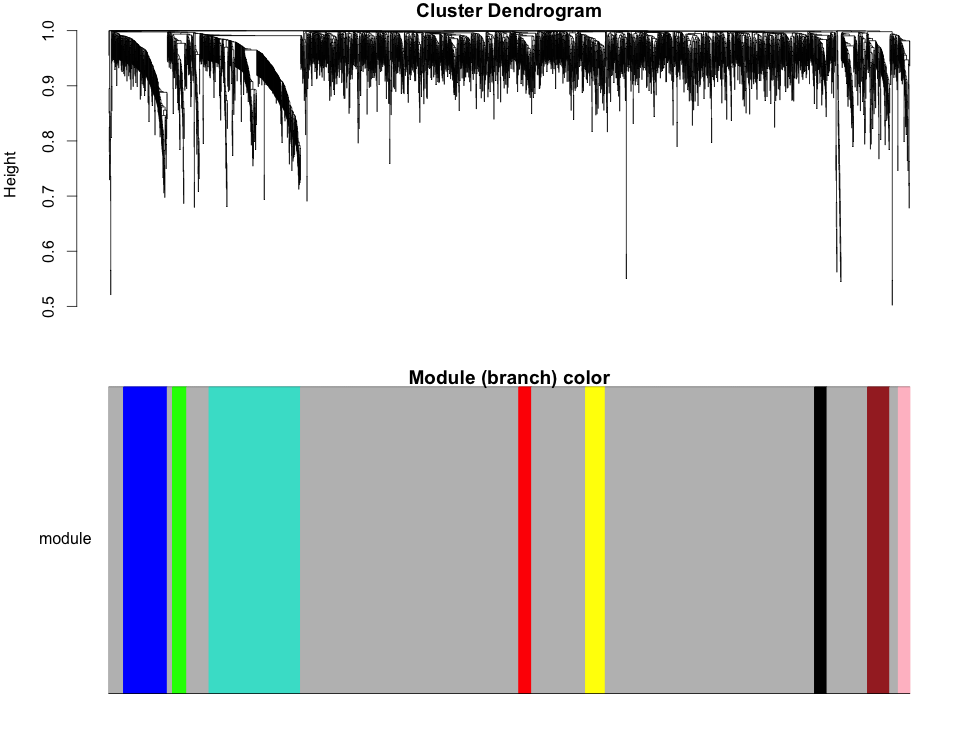}
\end{center}
\caption{Hierarchical Clustering and Module finding in Microarray dataset}
\end{figure}

To see how these modules were connected, we can create a clustering tree of the eigengenes for each module(figure 9).

\begin{figure}[h]
\begin{center}
\includegraphics[width=4in]{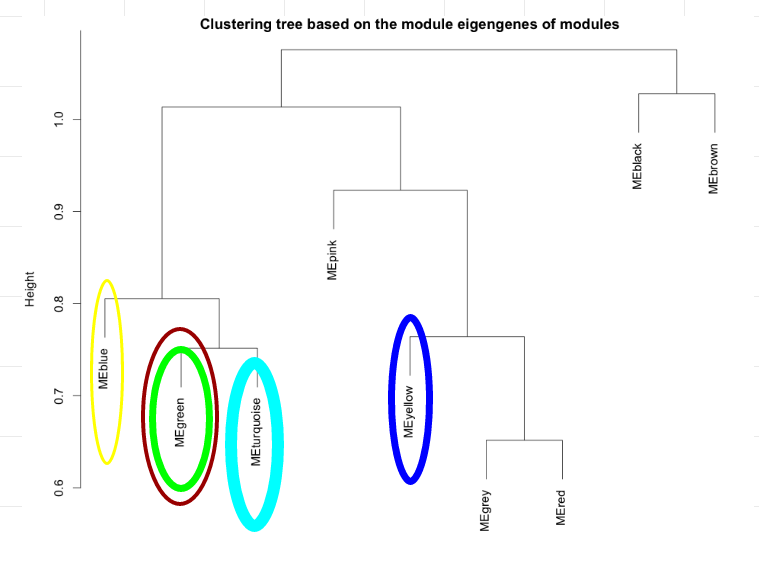}
\end{center}
\caption{Cluster tree for the Eigengenes in Microarray data}
\end{figure}

Many significant genes that were found in the microarray data set were found again in the RNA-seq data set. For each module here, we have circled it with the color of the RNA-seq module that has genes overlapped with the microarray module. For example, the blue module in the microarray data set had 51 significant genes from which 31 were found in the RNA-seq yellow module. From 85 genes in the turquoise module in array, 83 were found in the corresponding turquoise module in RNA-seq data. From 23 significant genes in the green module in the array data set, 22 were found in the corresponding green module in RNA-seq data. From 19 significant genes in the yellow module in array data, all 19 genes were found in the blue module in the RNA-seq data. These results increase our confidence in the original findings and their biological interpretations. 

\section{Conclusion}
At the beginning of the project, our goal was to do a thorough exploratory analysis of lung squamous carcinoma, and try to find genes and modules that are significant in terms of survival time as well as playing a crucial role in cancer growth. Our results gave us two genes which are highly connected as well as prognostically significant. The literature available on these two genes so far support our findings. Moreover, the highly connected genes in the significant modules are also shown to be involved in processes that are crucially connected to cancer cell growth. At the same time, the gene enrichment analysis showed some functions like involvement with glycoprotein or phosphoprotein where the immediate biological connection with lung cancer is not really clear. Biological experiments targeting these findings in the future might shed more light on these findings.

\textbf{This work was done as a final project for the BCB 570: Systems biology course (Spring 2016)}

\section{References}

\small{
[1]Langfelder, P. \& Horvath, S. (2008) {\it WGCNA: an R package for weighted correlation network analysis.} \url{http://bmcbioinformatics.biomedcentral.com/articles/10.1186/1471-2105-9-559} \textbf{The paper on the WGCNA R package.}

[2]\url{https://labs.genetics.ucla.edu/horvath/CoexpressionNetwork/Rpackages/WGCNA} \textbf{This website contains excellent tutorial on how to use WGCNA library to conduct an analysis.}

[3]{\it Should the Cox proportional hazards modelget the nobel prize?} \url{http://simplystatistics.org/2012/12/17/should-the-cox-proportional-hazards-model-get-the-nobel-prize-in-medicine/}

[4]Mischel, P. S. et al. {\it Analysis of oncogenic signaling networks in glioblastoma identifies ASPM as a molecular target.} \url{http://www.pnas.org/content/103/46/17402.full.pdf} \textbf{Similar work to my analysis but on Brain Cancer dataset.}

[5]UCSC Cancer genome browser.\url{https://genome-cancer.ucsc.edu/proj/site/hgHeatmap/} \textbf{Great website for many cancer datasets.}

[6]Justilien, V. {\it The guanine nucleotide exchange factor (GEF) Ect2 is an oncogene in human cancer.} \url{http://www.ncbi.nlm.nih.gov/pmc/articles/PMC2863999} \textbf{This paper shows the impact on lung cancer of ECT2 gene.}

[7]Fields, A. P. {\it Ect2 links the PKCiota-Par6alpha complex to Rac1 activation and cellular transformation.}

[8]Daigo Y. {\it Involvement of epithelial cell transforming sequence-2 oncoantigen in lung and esophageal cancer progression.}\url{http://www.ncbi.nlm.nih.gov/pubmed/19118053}

[9]Ragan, M. A. {\it Understanding the functional impact of copy number alterations in breast cancer using a network modeling approach.} \url{http://www.ncbi.nlm.nih.gov/pubmed/26805938}

[10]{\it DAVID - Gene enrichment analysis tool.} \url{https://david.ncifcrf.gov/home.jsp}

[11]Gibson, B.W. {\it Phosphoprotein Secretome of Tumor Cells as a Source of Candidates for Breast Cancer Biomarkers in Plasma.}\url{http://www.ncbi.nlm.nih.gov/pmc/articles/PMC3977182/}

[12]Zhang, B. \& Horvath, S. (2005) "A General Framework for Weighted Gene CoExpression
Network Analysis", Statistical Applications in Genetics and Molecular Biology:
Vol. 4: No. 1, Article 17. \url{https://labs.genetics.ucla.edu/horvath/GeneralFramework/GBMTutorialHorvath.pdf} \textbf{The excellent WGCNA brain cancer tutorial for which it was possible for me to do this analysis.}

\end{document}